\documentclass[twocolumn,prb,amsmath,amssymb,amsfonts,superscriptaddress,floatfix,nopacs]{revtex4}
\usepackage{graphicx}
\usepackage{dcolumn}
\usepackage{bm}
\usepackage{rotating}
\usepackage{color}

\def\be{\begin{equation}}
\def\ee{\end{equation}}
\def\bd{\begin{displaymath}}
\def\ed{\end{displaymath}}
\def\-{\phantom{-}}

\begin{document}

\title{Computational investigation of the electronic and optical properties of planar Ga-doped Graphene}

\date{\today}

\author{Nicole Creange}
\affiliation{Department of Physics and Astronomy, James Madison University, Harrisonburg, VA 22807, USA}
\author{Costel Constantin}
\affiliation{Department of Physics and Astronomy, James Madison University, Harrisonburg, VA 22807, USA}
\author{Jian-Xin Zhu}
\affiliation
{Theoretical Division, Los Alamos National Laboratory, Los Alamos, New Mexico 87545}
\affiliation
{Center for Integrated Nanotechnologies, Los Alamos National Laboratory, Los Alamos, New Mexico 87545}
\author{Alexander V. Balatsky}
\affiliation
{Institute of Material Science, Los Alamos National Laboratory, Los Alamos, New Mexico 87545}
\affiliation
{Nordic Institute for Theoretical Physics, KTH Royal Institute of Technology and Stockholm University, Roslagstullsbacken 23, 106 91 Stockholm, Sweden}

\author{Jason. T. Haraldsen}
\affiliation{Department of Physics and Astronomy, James Madison University, Harrisonburg, VA 22807, USA}

\begin{abstract}


We simulate the optical and electrical responses in gallium-doped graphene. Using density functional theory with a local density approximation, we simlutate the electronic band structure and show the effects of impurity doping (0-3.91\%) in graphene on the electron density, refractive index, optical conductivity, and extinction coefficient for each doping percentages.  Here, gallium atoms are placed randomly (using a 5-point average) throughout a 128-atom sheet of graphene. These calculations demonstrate the effects of hole doping due to direct atomic substitution, where it is found that a disruption in the electronic structure and electron density for small doping levels is due to impurity scattering of the electrons. However, the system continues to produce metallic or semi-metallic behavior with increasing doping levels. These calculations are compared to a purely theoretical 100\% Ga sheet for comparison of conductivity. Furthermore, we examine the change in the electronic band structure, where the introduction of gallium electronic bands produces a shift in the electron bands and dissolves the characteristic Dirac cone within graphene, which leads to better electron mobility.

\end{abstract}

\maketitle

\section{Introduction}

Understanding the complex properties of materials is a fundamental goal for the advancement of technologies and other industries in the world today. Specifically, the examination and prediction of electronic properties and phase transitions in insulators and metals has gained much attention, typically for the investigations of exotic topological states and/or large symmetry-breaking regimes\cite{wehl:14}. Topological materials in particular have been gaining a lot of attention due to their many surface properties. Specifically, the study of the electronic properties of graphene and its interactions with substrates and other materials has show to exhibit many desired phenomena for technological devices\cite{geim:07,neto:09}.

\begin{figure}
\includegraphics[width=3.25in]{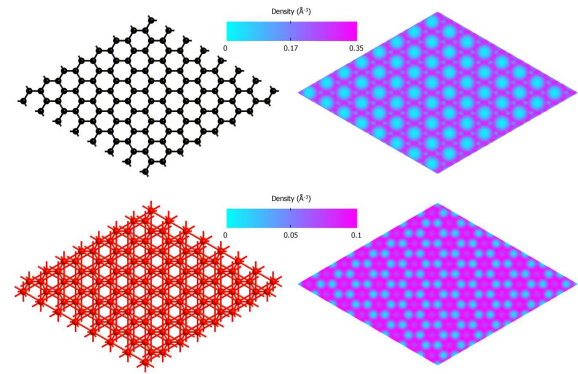}
\caption{The structure (left) and calculated energy density (right) for a pure graphene sheet (top) and theoretical gallium sheet (bottom). 
}
\label{strucED}
\end{figure}

Graphene is a sp$^2$-bonded 2D sheet of carbon atoms arranged in a honeycomb pattern (shown in Fig. \ref{strucED})\cite{zhu:10,novo:05,neto:09}. The bonding and arrangement of atoms allows graphene to have a very large tensile strength\cite{bolo:08} and interesting thermal properties\cite{pop:12,bala:08}, which when considering its size and weight, is stronger than steel\cite{sava:12}. Furthermore, the $\pi$-bonding throughout the lattice provides a high electron mobility\cite{lee:08,moro:08}, where the high durability, strength, and conductivity make graphene a suitable choice for use in nanodevices and electronics. However, even though graphene has a high tensile strength, the ability for the material to be a great component for electronics is still unclear.

\begin{figure*}
\includegraphics[width=6.25in]{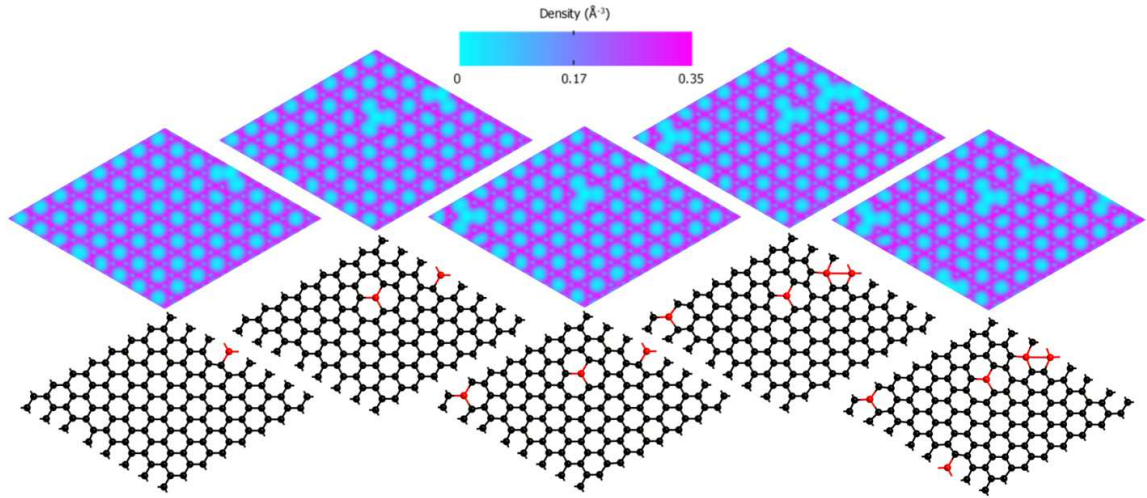}
\caption{The calculated electron density as a function of spatial position in graphene with various concentrations of gallium (from left to right - 0.78, 1.56, 2.34, 3.13, and 3.90\% Ga doping). The crystal structure is given directly below the density plots.
}
\label{ED}
\end{figure*}

Currently, Graphene is being investigated for use in electronics through surface deposition on metals and semiconductors, where the interactions of graphene and other atoms is hoped to produce enhancements to the overall electronic properties\cite{chen:08}. This enhancement is due to the presence of surface or topological states\cite{wehl:14}, which are critical to the understanding of the electron mobility in graphene, and has led to many researchers working to find ways for improving the electronic structure and overall electronic properties of graphene. While topological interactions with substrates and other materials plays an important role these advancements, the ability to change and control the electronic properties of graphene through direct substitution could provide a unique tunability to the mechanisms that drive technology further. 


Recent experiments have demonstrated that it is possible to remove individual carbon atoms from the graphene lattice\cite{gome:12}. Furthermore, various experimental and theoretical papers have also shown that carbon atoms can be replaced with other atoms (i.e. nitrogen, cobalt, and iron) in specifically designed spots\cite{robe:13,sant:10,hu:11,wang:12,lv:12}. This opens a whole realm of possibilities for the tuning of the electronic properties of 2D materials through a manipulation of the electronic and topological states.

\begin{figure*}
\includegraphics[width=8.0in]{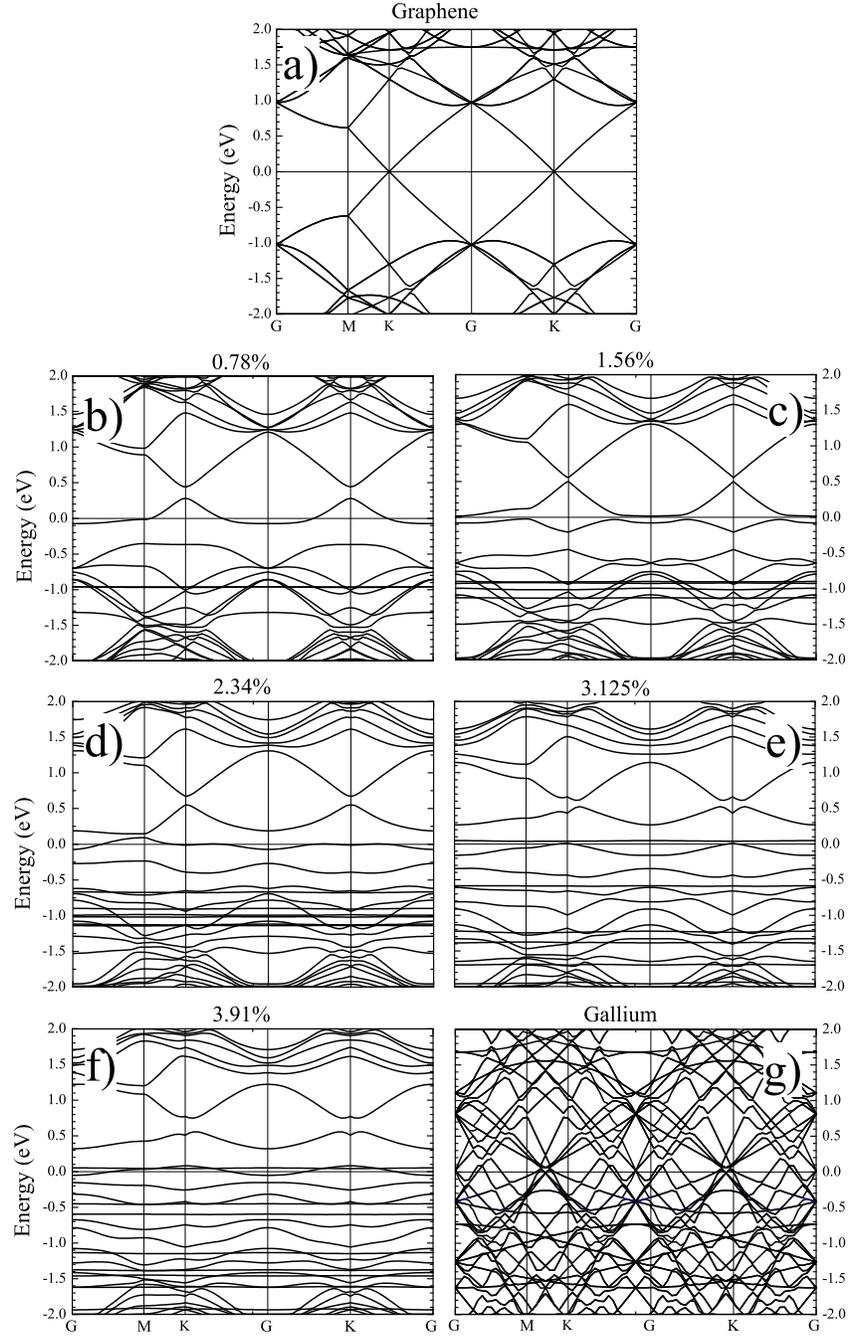}
\caption{The simulated electronic band structure for graphene sheets with 0\%(a), 0.78\%(b), 1.56\%(c), 2.34\%(d), 3.13\%(e), 3.91\%(f), and 100\%(g) substitution of gallium.
}
\label{BS}
\end{figure*}

Topological states in graphene are produced through a unique conductivity point in its band structure known as the Dirac point, which typically allows electrons to easily move between the conduction and valence bands through a band structure Dirac cone at the Fermi level. The introduction of substituted or doped atoms will produce a breaking of the Dirac cone. However, the presence of these atoms will produce an impurity resonance at the Fermi level, which should provide an increase in the electron mobility and overall conductance. In many case thus far, researchers have examined electron favorable or magnetic atoms. Therefore, we propose the idea that impurity states may be hindered through hole doping or p-type substitution in graphene using gallium atoms.

To examine the effects of topological states produced by p-substituted atoms, we examine the presence of directly-substituted Gallium atoms in graphene at low doping levels of 0 - 3.91\%. Using density functional theory in an local density approximation (LDA), we determine the electronic structure and optical properties for the various stages of Ga doping. Overall, we find that low doping of graphene with Ga will actually produce a decrease in the overall conductivity, which is a trend observed in the index of refraction and extinction coefficient as well. To reconcile these observations with known production of impurity resonances in Dirac materials, we examine the electron density and compare calculations to a theoretical 2D structure of pure Ga atoms. Here, we determine that the presence of a small amount p-type dopant produces impurity scattering in the graphene sheet, which is the likely cause for the decrease in overall electron mobility. 

\begin{figure}
\includegraphics[width=3.50in]{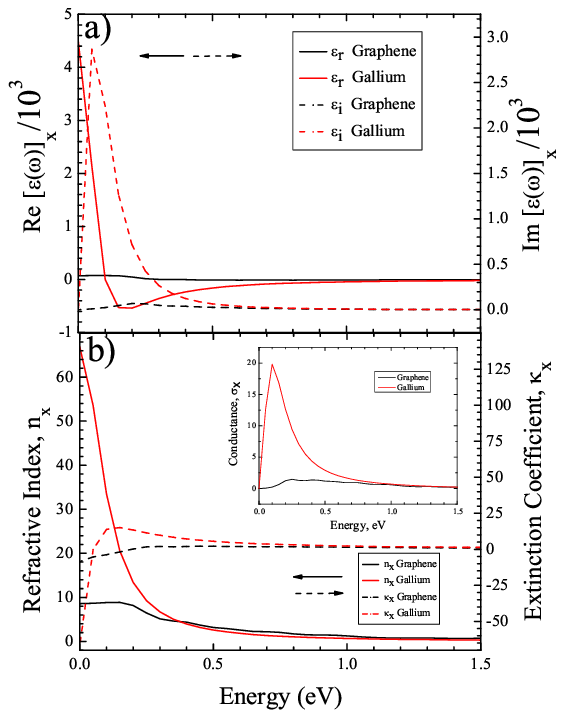}
\caption{Simulated dielectric response as a function of energy for a pure graphene sheet and a theoretical gallium sheet. The bottom panel shows the calculated refractive index and extinction coefficient as functions of energy, where the conductance is placed in the inset.
}
\label{OS-cond}
\end{figure}

\section{Computational Methodology}

We ran a series of density functional calculations using Atomistix Toolkit by Quantumwise\cite{quantumwise,bran:02,sole:02}. In this study, constructed a 128 atom supercell of graphene and subsequently replace random carbon atoms with Ga using doping levels of 0, 0.78, 1.56, 2.34, 3.125, 3.91, and 100\%. This is illustrated in Fig. \ref{ED}. To simulate unbiased placement, Ga atom positions were determined by a random number generator, and for systems with 2-5 impurity atoms, five different configurations were used to find the average quantities. 

Using a local density approximation (LDA), a geometry optimization and band structure calculations were performed using a k-point sampling of 5x5x1 with periodic boundary conditions for 23 different configurations. Furthermore, we determined the optical spectrum, and electron density and calculated the conductivity, refractive index, and extinction coefficient. The doped graphene structure was kept in the planar configuration to allow for us to examine only the electronic effects of substitution without energy parameters provided by lattice distortions. Further studies will analyze how the energy bands, conductivity, and optical properties are affected in the presence of a lattice distortion. The geometry optimization did produce a planar distortion due to the presence of the larger Ga atoms.

From the states in the electronic band structure and electron density\cite{grif:99}, the Kubo-Greenwood formula can be used to calculate the susceptibility tensor from state $m$ to $n$ is

\begin{equation}
\begin{array}{l}
\displaystyle \chi_{ij}(\omega) = - \frac{ e^2 \hbar^4}{m^2 \epsilon_0 V \omega^2}\\\\ 
\displaystyle \sum_{nm} \frac{f(E_m)-f(E_n)}{E_{n}-E_{m}-\hbar(\omega - i \eta)}
<n|j_{\alpha}|m><m|j_{\beta}|n>, 
\end{array}
\end{equation}

\noindent where $j_{\alpha}$ describes the momentum operator between matrix element between state $n$ and $m$, $V$ is the volume, $\eta$ is a finite broadening, and $f(E_m) = 1 / ({\rm exp}(E/k_B T) + 1)$\cite{alle:06}. From the susceptibility tensor\cite{mart:04}, we determine the real and imaginary parts of the dielectric tensor

\begin{equation}
\epsilon_r(\omega)  =  (1 + \chi(\omega) )
\end{equation}
and the optical conductivity
\begin{equation}
\sigma(\omega)  =  - i \omega \epsilon_0 \chi(\omega).
\end{equation}
From the dielectric tensor\cite{harr:70}, the refractive index and extinction coefficient can be calculated as
\begin{equation}
n  = \sqrt{\frac{\sqrt{\epsilon_1^2+\epsilon_2^2}+\epsilon_1}{2}}
\end{equation}
and the extinction coefficient
\begin{equation}
\kappa  =  \sqrt{ \frac{ \sqrt{ \epsilon_1^2+ \epsilon_2^2}-\epsilon_1}{2}}.
\end{equation}

Using these quantities, it is possible to evaluate the electron mobility of supercell.

\section{Results and Discussion}

Figure \ref{BS} shows the calculated electronic band structure for each configuration. This illustrates the dramatic change in electron mobility as Ga atoms are substituted into the graphene lattice. Once one Ga atom is substituted into graphene (Fig. \ref{BS}(b)), the characteristic Dirac cone shifts in energy and is broken and produces a distinct bands at the Fermi level. As more atoms are substituted into the 2D lattice, the impurity bands form at zero energy which produces the observed impurity resonances.  

From the calculated band structure, the dielectric tensor can be determined as a function of energy. Figure \ref{OS-cond}(a) shows the determined optical spectra for pure graphene and a theoretical analog of Ga atoms. This illustrates the expected increase in conductivity, which is characteristic of the simulated band structure.

\begin{figure}
\includegraphics[width=3.25in]{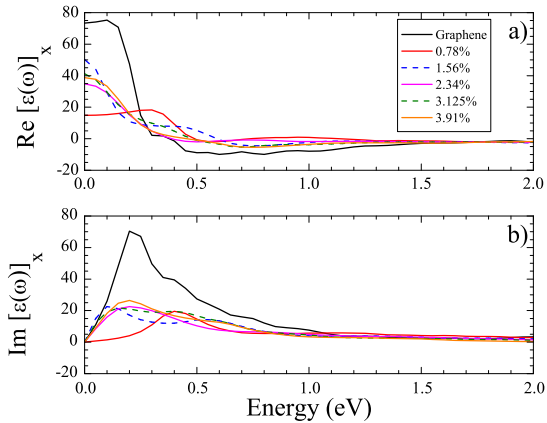}
\caption{Simulated $x$-axis dielectric response for the various concentrations of gallium (0, 0.78, 1.56, 2.34, 3.13, and 3.91\%). This is the average dielectric response consisting of a five-point configuration average. The average dielectric response along the $y$ axis is similar.  
}
\label{OS}
\end{figure}

\begin{figure}
\includegraphics[width=3.250in]{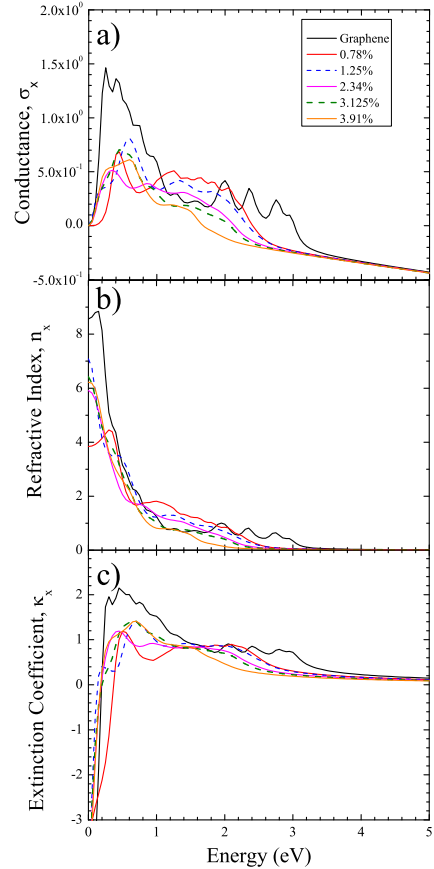}
\caption{The calculated conductance(a), refractive index (b), extinction coefficient (c) as function of energy for the various substitutions concentrations of gallium (0, 0.78, 1.56, 2.34, 3.13, and 3.91\%).
}
\label{cond-ga}
\end{figure}

Figure \ref{OS} shows the real and imaginary components of the calculated dielectric tensors for of doping cases. The dielectric response shows a low energy shift as Ga is added to the system. From the calculated optical spectra, we can determine the conductivity, refractive index, and extinction coefficient (shown in Fig. \ref{cond-ga}).

From the comparison of the optical spectra properties for graphene and low doping cases of Ga, there is a distinct lose of conductivity and electron mobility, which is in contradiction to the change in the optical properties from graphene to a pure Ga sheet. The optical properties also appear to be in slight contradiction to the electronic structure for the low doping Ga cases, where the impurity bands seem to provide cross-over and conduction states at the Fermi level.

To understand these contradictions, we examined the electron density (shown in Figs. \ref{strucED} and \ref{ED}) correlating to each configuration.  From the electron density in the graphene sheet, it is shown that the electrons are distributed fairly evenly in between each carbon atom in the pure graphene sheet, where blue regions represents no or a low electron density and red correlated to high electron density. However, when the first dopant is added, the electron density shifts away from this region, which is indicated by the large blue region. This shift indicated the movement of the electrons from the gallium atoms. This trend continues through the low concentration Ga-doped sheets. However, in the theoretical pure gallium sheet, the electrons tend to cluster right on top of each gallium atom rather than in between the atoms, which produces the large increase in the electronic mobility.

As we have seen in both the optical spectrum and electron configurations, there is a general disturbance in the system causing a decrease in electron mobility. From the electron density, this decrease in conductivity seems to be due to impurity scattering at the low doping levels, which will restrict the ability to enhance the electronic properties through direct substitution. The formation of low electron density holes from the Ga atoms produce transmission pathway disruptions, and as a gallium atom is added to the system, electrons encounters the holes in the graphene sheet and is scattered from them. This trend continues for all low doping percentages and the only change seems to occur at the pure gallium sheet. In this theoretical gallium sheet, the optical spectrum shows an extremely high conductance due to the disappearance of the impurity scattering from the uniformity of the sheet. 
The low to high conductivity from the doping cases to the theoretical Ga sheet indicates there must be a cross-over point in the intermediate region of doping. However, when that region was calculated the 2D sheet was found to be unstable in the geometry optimization. Therefore, it is only possible to examine the low-doping regions to understand the possibility of enhanced electronic and optical properties.

\section{Conclusions}

In this study, we have investigated the effect of low percentage gallium doping on graphene in order to understand the possibility of electronic enhancement.  Using density functional theory and a local density approximatione, it was found the as Ga dopants are added to the system, there is a distinct shift in the optical spectrum that indicates a decrease in the conductance. Through an analysis of the electron for all configurations, it appears this is due to impurity scattering at the Ga atomic site. This analysis was done in the confines of the planar geometry to minimize contributions produced by geometric distortions. Further work will incorporate the $z$-axis graphene distortion to provide a better understanding of the system.

\section*{Acknowledgements}

N.C., C.C., and J.T.H thank the support of the Institute for Materials Science at Los Alamos National Laboratory. The work at Los Alamos National Laboratory was carried out, in part, under the auspice of the U.S. DOE and NNSA under Contract No. DEAC52-06NA25396 and supported by U.S. DOE Basic Energy Sciences Office (A.V.B.). Work of A.V.B. was also supported by European Research Council (ERC) DM 321031. This work was also, in part, supported by the Center for Integrated Nanotechnologies, a U.S. DOE Office of Basic Energy Sciences user facility (J.-X.Z).

\end{document}